\documentclass[showpacs,showkeys,preprintnumbers]{revtex4}%
\usepackage{amsfonts}
\usepackage{amsmath}
\usepackage{graphicx}
\usepackage{dcolumn}
\usepackage{bm}
\usepackage{amssymb}%
\begin{document}
\title{Decay modes of the excited pseudoscalar glueball}
\author{Walaa I.\ Eshraim and Stefan Schramm}
\affiliation{Frankfurt Institute for Advanced Studies, Goethe University,
Ruth-Moufang-Str.\ 1, D--60438 Frankfurt am Main, Germany }

\begin{abstract}
We study three different chiral Lagrangians that describe the two- and three-body decays of an excited pseudoscalar glueball, $J^{PC}=0^{*-+}$, into light mesons and charmonium states as well as into a scalar and pseudoscalar glueball. We compute the decay channels for an excited pseudoscalar glueball with a mass of $3.7$ GeV and consider a ground state pseudoscalar glueball of mass $2.6$ GeV, following predictions from lattice QCD simulations. These states and channels are in reach of the ongoing BESIII experiment and the PANDA experiments at the upcoming FAIR facility experiment. We present the resulting decay branching ratios with a parameter-free prediction.

\end{abstract}

\pacs{12.39.Fe, 12.39.Mk, 13.20.Jf}
\keywords{chiral Lagrangians, (pseudo-) scalar mesons, pseudoscalar glueball}\maketitle

\section{Introduction}

Glueballs, the bound states of gluons, form colorless, or 'white', states, predicted by Quantum Chromodynamic (QCD) \cite{bag-glueball}, the theory of strong interactions. The fundamental symmetry of QCD is the exact local $SU(3)_c$ color symmetry and, due to the non-Abelian nature \cite{Nakano} of this symmetry, the gauge fields interact with each other strongly. This interaction gives rise to a color-singlet state, which consists of gluons, the so-called glueball. Considering the quarks as well, the glueball will be a mixed state of gluons and  ($\overline{q}q$)-meson states with the same spin and parity.\\
\indent The investigation of the properties of glueballs is an important field in hadronic physics and has been extensively studied, starting with the computation of the glueball mass using the bag-model \cite{bag-glueball} as well as in the flux-tube model. The glueball spectrum was also computed via lattice simulations of Yang-Mills theory \cite{general, pseudo, expseudo}. Note that in full QCD (i.e., gluons plus quarks), the mixing of glueball and quark-antiquark configurations with the same quantum number occurs  complicating the identification of the corresponding resonances  as listed in the Particle Data Group (PDG) \cite{PDG}. The experimental and theoretical efforts (see Refs. \cite{review, staniglueball, EshraimG, EshraimTH} and refs. therein) in searching for (predominantly) glueball states represent important steps towards a better understanding of the non-perturbative behaviour of QCD. However, this search entails the complex task of identifying glueballs unambiguously. Generally, there are two key properties assisting with determining a glueball state through its decays: these should be narrow and exhibit 'flavour blindness'. However, one has found an exception in the decays of the scalar glueball $f_0(1710)$, which preferentially decays into kaons and $\eta$ mesons and less into pions, in contrast to the 'flavour blindness' condition. This peculiar result for the $f_0(1710)$ decays has been attributed to a 'chiral suppression' mechanism \cite{Carlson, Sexton, Chanowitz} according to which the decay amplitudes of the glueball is proportional to the current quark mass in the final state.\\
\indent The numerical approach of lattice QCD has been employed extensively to compute the glueball spectrum \cite{general, pseudo, expseudo, Lattice}, where the lightest glueball state has been found to be a scalar-isoscalar state, $J^{PC}=0^{++}$ , with a mass of about $1.7$ GeV. This energy region has been studied in a variety of effective approaches \cite{scalars, stani, Lee, FGf0}. As a result, the measured resonance $f_0(1710)$ appears to be a glueball candidate for several reasons: firstly, its mass is very close to that of the lattice QCD value, and secondly, its properties fit the phenomenology of the scalar glueball as calculated in the extended Linear Sigma Model (eLSM) in Ref.  \cite{staniglueball},  the phenomenological solutions as seen in Ref. \cite {FGf0}, the Lattice study in Refs. \cite{Lee, Gui}, and the combination of Lattice QCD calculations and experimental data for disentangling the glue and $\overline{q}q$ components of the scalar glueball in Ref. \cite{Cheng}. Lastly, it is profusely produced in the gluon-rich decay of the $J/\psi$ meson. The second lightest glueball state has been predicted with a tensor quantum number ($J^{PC}=2^{++}$) and a mass of about $2.2$ GeV. The resonance $f_J(2200)$ could be a very good candidate \cite{tensor, burakovsky}, in the case that its total spin is experimentally confirmed to be $J=2$. \\
\indent The third lightest glueball predicted by lattice QCD simulation is a pseudoscalar glueball ($J^{PC}=0^{-+}$) with a mass of about $2.6$ GeV \cite{pseudo, expseudo}. The range of the mass of the pseudoscalar glueball has been predicted to vary from the $\eta(1405)$ (or $\iota(1440$)) to $2.6$ GeV. Moreover, the state $X(1835)$ has been investigated as a pseudoscalar glueball by using an effective Lagrangian approach \cite{Li}. Beside that the two states $X(2120)$ and $X(2370)$ have been interpreted as a glueball in Ref. \cite{Yu}. In Ref.\cite{EshraimG} we studied the decay properties of the lightest pseudoscalar glueball within the eLSM in the case of three flavours in two scenarios: the first assuming the mass of the pseudoscalar glueball to be in agreement with lattice QCD, and the second scenario where the pseudoscalar glueball has a mass slightly lower than the lattice QCD prediction. This is motivated by the BESIII experiment, where pseudoscalar states have been investigated in $J/\psi$ decays \cite{bes} with a measured resonance $X(2370)$ with a mass of $2.37$ GeV as promising potential glueball candidate. Furthermore, in our study of pseudoscalar glueballs, we include the first two states ($J^{PC}=0^{-+}$) as determined in lattice QCD. Here, in quenched approximation \cite{expseudo} the first excited $0^{*-+}$ state has a mass of $3.7$ GeV, which will be included in our investigation.\\

\indent In this work we study the decay properties of the first excited pseudoscalar glueball state whose mass lie, in agreement with lattice QCD, between $3$ and $4$ GeV. We constructed three  effective Lagrangian: (i) The first involves interaction of the excited pseudoscalar glueball $\tilde{G}$ with the lowest pseudoscalar glueball $\tilde{G}'$ and (pseudo)scalar mesons in the three-flavour case. We can thus evaluate the widths for the decays $\Gamma_{\tilde{G} \rightarrow \tilde{G}' PP}$, where $P$ refers to pseudoscalar quark-antiquark states which are the well-known light pseudoscalars $\{\pi,\,K,\,\eta,\,\eta'\}$ fixing the mass of the pseudoscalar glueball from lattice QCD at $2.6$ GeV. (ii) The second effective Lagrangian couples the excited pseudoscalar glueball $\tilde{G}$ with a scalar glueball $G$ and (pseudo)scalar mesons in the $N_f=3$ case. Accordingly, we can compute the two- and three-body decay widths of the pseudoscalar glueball into (pseudo)scalar mesons, where the quark-antiquark nonet of scalars is above $1$ GeV: $\{a_0(1450),\, K_S,\, f_0(1370),\,f_0(1500),\,f_0(1710)\}$, and scalar glueball, which corresponds to the resonance $f_0(1710)$ as discussed in Ref. \cite{staniglueball} and/ or admixtures of the resonances $f_0(1500)$ and $f_0(1710)$.\\
(iii) The third Lagrangian term couples the excited pseudoscalar glueball with the scalar and pseudoscalar mesons in the case of four-flavours (that is, including charmed mesons) \cite{EshraimC}. This allows us to calculate the decay of the first excited pseudoscalar glueball into the charmonium state $\eta_C$, as $\Gamma_{\tilde{G} \rightarrow \eta_C\pi\pi}$,  and the two- and three-body decay widths including (pseudo)scalar mesons with the same channels produced by the second effective Lagrangian. Note that the charmonium state $\eta_C$ could decay into the pseudoscalar glueball $\tilde{G}'$, as $\Gamma_{\eta_C \rightarrow \tilde{G}' \pi\pi}$, as seen in Refs. \cite{EshraimTH, WalICNFP15}. \\
The three chiral Lagrangians that we consider involve three unknown coupling constants, which cannot be fixed without experimental data. Therefore, we compute the branching ratios and present a useful guideline for experimental investigations into the excited pseudoscalar glueball state. This is of particular relevance for the upcoming PANDA experiment at the FAIR facility \cite{panda}, for the BESIII experiment \cite{bes} and for NICA \cite{DenisNICA}, which has the ability to measure the proposed channels. PANDA  will use an 1.5 GeV antiproton beam on a proton target  at rest, yielding sufficient energy to produce directly the excited pseudoscalar glueball as an intermediate state. NICA will study charmonium systems, which also allows for reconstructing potential glueball states.\\
\indent This paper is organized as follows. In Sec. II we present the effective Lagrangian interaction between the excited pseudoscalar glueball and the pseudoscalar glueball as well as scalar, and pseudoscalar quark-antiquark degrees of freedom, allowing for the computation of the branching ratios for the decays into $\tilde{G}'PP$. In Sec. III we present a chiral Lagrangian term that couples the pseudoscalar glueball with the scalar glueball, scalar, and pseudoscalar mesons in the three-flavour case, as well as an extended chiral Lagrangian connecting the excited pseudoscalar glueball to the (pseudo)scalar mesons in the case of $N_f=4$. With this approach, we evaluate the branching ratios for the decays into two- and three-body. Finally, in Sec. IV we present our conclusions.

\section{Decay of an excited pseudoscalar glueball into the lowest pseudoscalar glueball}

We  introduce a chiral Lagrangian which couples the excited pseudoscalar glueball $\tilde{G}\equiv\left\vert gg\right\rangle$
with quantum numbers $J^{PC}=0^{-+}$ to a pseudoscalar glueball $\tilde{G}'\equiv\left\vert gg\right\rangle$ with the same quantum number and to the ordinary scalar and pseudoscalar mesons.
\begin{equation}
\mathcal{L}_{\tilde{G}\tilde{G}'}^{int}=c_{\tilde{G}\tilde{G}'}\tilde{G}\tilde{G}'\,Tr\left(\Phi^\dag\Phi\right)  \text{ ,}
\label{intlag}%
\end{equation}
where $c_{\tilde{G}\tilde{G}'}$ is a coupling constant, and 
\begin{equation}
\Phi=(S^{a}+iP^{a})t^{a} \label{phimat}%
\end{equation}
is a multiplet containing the usual scalar and pseudoscalar quark-antiquark states. The $t^a$ are the generators of the group $U(N_f)$. In the Lagrangian (\ref{intlag}), we consider the case $N_f=3$, thus $c_{\tilde{G}\tilde{G}'}$ is dimensionless, and the explicit representation of the scalar and pseudoscalar mesons reads \cite{dick}:%
\begin{equation}
\Phi=\frac{1}{\sqrt{2}}\left(
\begin{array}
[c]{ccc}%
\frac{(\sigma_{N}+a_{0}^{0})+i(\eta_{N}+\pi^{0})}{\sqrt{2}} & a_{0}^{+}%
+i\pi^{+} & K_{S}^{+}+iK^{+}\\
a_{0}^{-}+i\pi^{-} & \frac{(\sigma_{N}-a_{0}^{0})+i(\eta_{N}-\pi^{0})}%
{\sqrt{2}} & K_{S}^{0}+iK^{0}\\
K_{S}^{-}+iK^{-} & \bar{K}_{S}^{0}+i\bar{K}^{0} & \sigma_{S}+i\eta_{S}%
\end{array}
\right)  \; . \label{phimatex}%
\end{equation}
which transforms as $\Phi\rightarrow U_{L}\Phi U_{R}^{\dagger}$ under $U_{L}(3)\times U_{R}(3)$  chiral transformation where $U_{L(R)}=e^{-i\Theta^a_L(R)^{t^a}}$ are $U(3)_{L(R)}$ matrices. Performing these transformations on the determinant of the multiplet $\Phi$, we see that this object is invariant under $SU_{L}(3)\times SU_{R}(3)$, but not under the axial $U(1)_A$ transformation.
\begin{equation}
det \Phi \rightarrow det U_A\Phi U_A=e^{-i\Theta^0_A\sqrt{2N_f}}det\Phi\neq det\Phi,
\end{equation} 
which is in agreement with the chiral anomaly. Moreover, the pseudoscalar glueball field $\tilde{G}$ and the excited pseudoscalar field $\tilde{G}'$ are invariant under $U(3)_L\times U(3)_R$ transformations. In addition, the pseudoscalar glueball, the excited pseudoscalar glueball and the quark-antiquark multiplet transform under the charge conjugation $C$ and the parity $P$ as 
$$\tilde{G}'(t,\overrightarrow{x})\rightarrow - \tilde{G}'(t,\overrightarrow{x}),\,\,\,\,\,\tilde{G}(t,\overrightarrow{x})\rightarrow - \tilde{G}(t,\overrightarrow{x}),\,\,\,\,\,\Phi(t,\overrightarrow{x})\rightarrow \Phi^\dag(t,\overrightarrow{x})\,,$$
and under charge conjugation as 
$$\tilde{G}'\rightarrow \tilde{G}',\,\,\,\,\,\tilde{G}\rightarrow \tilde{G},\,\,\,\,\,\Phi\rightarrow \Phi^T\,.$$
Consequently the effective chiral Lagrangian (\ref{intlag}) possesses the symmetries of the QCD Lagrangian, which is invariant under $SU(3)_R\times SU(3)_L$ symmetry, parity and charge conjugate but it is not invariant with respect to the axial $U(1)_A$ following the axial anomaly in the isoscalar-pseudoscalar sector.\\
The states in Eq.(2) are assigned as physical resonances to light quark-antiquark states with mass $\lesssim 2$ GeV \cite{dick} as follows: (i) In the pseudoscalar sector $P$, the fields $\overrightarrow{\pi}$ and $K$ represent the pion isotriplet and the kaon isodoublet respectively \cite{PDG}. The bare quark-antiquark fields $\eta_{N}%
\equiv\left\vert \bar{u}u+\bar{d}d\right\rangle /\sqrt{2}$ and $\eta_{S}%
\equiv\left\vert \bar{s}s\right\rangle $ are the non-strange and strangeness mixing
components of the physical states $\eta$ and $\eta^{\prime}$ which can be obtained by \cite{PDG}:%
\begin{equation}
\eta=\eta_{N}\cos\varphi+\eta_{S}\sin\varphi,\text{ }\eta^{\prime}=-\eta
_{N}\sin\varphi+\eta_{S}\cos\varphi, \label{mixetas}%
\end{equation}
where the mixing angle is $\varphi\simeq-44.6^{\circ}$ \cite{dick}. There are two different values for the mixing angle, e.g. $\varphi=-36^°$ or $\varphi=-41.4^°$, determined by the KLOE Collaboration \cite{KLOEpsglue} but this uncertainty has only a minor effect on the present investigation. (ii) In the scalar sector $S$, the field $\vec{a}_{0}$ corresponds to the
physical isotriplet state $a_{0}(1450)$ and the scalar kaon field $K_{S}$ is identified with
the physical isodoublet state $K_{0}^{\star}(1430).$ In the scalar-isoscalar sector, the non-strange bare field $\sigma_{N}\equiv\left\vert \bar{u}u+\bar{d}d\right\rangle /\sqrt{2}$
corresponds to the resonance $f_{0}(1370)$ and the bare strange field $\sigma_{S}$ corresponds to $f_0(1500)$ \cite{staniglueball}, which mixes with the scalar glueball, G, with amixing matrix as constructed in Ref.\cite{staniglueball}:
\begin{equation}\label{scalmixmat}
\left(%
\begin{array}{c}
 f_0(1370) \\
 f_0(1500)   \\
 f_0(1710)\\
\end{array}%
\right)=\left(%
\begin{array}{ccccc}
 -0.91 & 0.24 & -0.33\\
 0.30 & 0.94  &-0.17\\
 -0.27 & 0.26 & 0.94\\
\end{array}%
\right)\left(%
\begin{array}{c}
 \sigma_N\\
  \sigma_S\\
  G\\
\end{array}%
\right).
\end{equation}
which gives
\begin{align}
 \sigma_N &=0.94 f_0(1370)+0.21f_0(1500)-0.26f_0(1710)\,,\\
 \sigma_S &=-0.17f_0(1370)+0.97f_0(1500)+0.18f_0(1710)\,,\\
  G &=-0.33f_0(1370)-0.172f_0(1500)+0.93f_0(1710)\,.\label{Gscalar}
\end{align}

To evaluate the decays of the excited pseudoscalar glueball $\tilde{G}$ we have to implement the effect of spontaneous symmetry breaking by shifting the scalar-isoscalar fields by their vacuum expectation values as follows \cite{dick} 

\begin{equation}
\sigma_{N}\rightarrow\sigma_{N}+\phi_{N}\text{ and }\sigma_{S}\rightarrow
\sigma_{S}+\phi_{S}\text{ .} \label{shift}%
\end{equation}

where $\phi_N$ and $\phi_S$ are the corresponding chiral condensates, which read

\begin{align}
\phi_{N}=&Z_{\pi}f_{\pi}=0.158\text{ GeV, }\\ \nonumber
\phi_{S}=&\frac{2Z_{K}f_{K}-\phi_{N}%
}{\sqrt{2}}=0.138\text{ GeV}\;,
\end{align}
where the value of the decay constant of the pion is $f_{\pi}=0.0922$ GeV, while the kaon decay constant  is given as $f_{K}=0.110$ \cite{PDG}. In order for the (axial-)vector mesons to appear in the Lagrangian (\ref{intlag}), one has also to consider the shifting of the axial-vector fields and thus to redefine the wave function of the pseudoscalar fields 
\begin{equation}
\vec{\pi}\rightarrow Z_{\pi}\vec{\pi}\text{ , }K^{i}\rightarrow Z_{K}%
K^{i}\text{, }\eta_{j}\rightarrow Z_{\eta_{j}}\eta_{j}\;, \label{psz}%
\end{equation}
whereas $i=1,2,3$ refers to the four kaonic fields. The
numerical values of the renormalization constants of the corresponding wave functions are $Z_{\pi}=1.709$,
$Z_{K}=1.604,Z_{K_{S}}=1.001,$ $Z_{\eta_{N}}=Z_{\pi},$ $Z_{\eta_{S}}=1.539$ \cite{dick}. By using Eqs.\ (\ref{shift}) and (\ref{psz}), the Lagrangian in Eq.\ (\ref{intlag}) includes the relevant tree-level vertices for the decay processes of $\tilde{G}$, see
Appendix (Sec.\ \ref{app2}).\\
Now we can determine the branching ratios of the excited pseudoscalar glueball, $\tilde{G}$, for the three-body decay into a pseudoscalar glueball $\tilde{G}'$ and two pseudoscalar mesons ($\Gamma_{\tilde{G} \rightarrow \tilde{G}' PP}$).  
We present the branching ratios relative to the total decay width of the pseudoscalar glueball $\Gamma_G^{tot}$. (The details of the calculation of the three-body decay is given in Appendix A5.)\\

\begin{center}%
\begin{table}[h] \centering
\begin{tabular}
[c]{|c|c|c|}\hline
Quantity & The theoretical result \\\hline
$\Gamma_{\tilde{G}\rightarrow \tilde{G}'KK}/\Gamma_{\tilde{G}}^{tot}$ & $0.0277$ \\\hline
$\Gamma_{\tilde{G}\rightarrow \tilde{G}'\pi\pi}/\Gamma_{\tilde{G}}^{tot}$ & $0.9697$\\\hline
$\Gamma_{\tilde{G}\rightarrow \tilde{G}' \eta\eta'}/\Gamma_{\tilde{G}}^{tot}$ & $0.0026$\\\hline
$\Gamma_{\tilde{G}\rightarrow  \tilde{G}'\eta\eta}/\Gamma_{\tilde{G}}^{tot}$ & $0.000012$ \\\hline

\end{tabular}%
\caption{Branching ratios for the decay of the excited pseudoscalar glueball $\tilde
{G}$ into the pseudoscalar glueball $\tilde{G}'$.}%
\end{table}%
\end{center}

Note that the results are presented as branching ratios because of the undetermined coupling constant $c_{\tilde{G}\tilde{G}'}$. The three body decay mode $\Gamma_{\tilde{G}\rightarrow \tilde{G}'\pi\pi}$ almost saturates the decay channels due to the small mass of the pions.

\section{Decay of an excited pseudoscalar glueball into scalar-isoscalar, (pseudo)scalar, and charmonium states}

We consider a chiral Lagrangian that couples the excited pseudoscalar glueball and a scalar glueball $G\equiv\left\vert gg\right\rangle$ with quantum number $J^{PC}=0^{-+}$ to scalar and pseudoscalar mesons.

\begin{equation}
\mathcal{L}_{\tilde{G}G}^{int}=ic_{\tilde{G}G\Phi}\tilde{G}G\left(
\text{\textrm{det}}\Phi-\text{\textrm{det}}\Phi^{\dag}\right)  \text{ ,}
\label{intlag2}%
\end{equation}

where $c_{\tilde{G}G\Phi}$ is an unknown coupling constant and $\Phi$ is a multiplet of a scalar and a pseudoscalar glueball in the case of $N_f=3$ as shown in Eq.(\ref{phimatex}). The effective Lagrangian of Eq.(\ref{intlag2}) is invariant under $SU_{L}(3)\times SU_{R}(3)$ and  parity. Applying the mixing matrix (\ref{scalmixmat}), the scalar glueball $G$ corresponds to the resonance $f_0(1710)$ \cite{staniglueball} as seen in Eq.(\ref{Gscalar}).\\
One has to perform the field transformations in Eq.(\ref{shift}) and Eq.(\ref{psz}) as well as shift the scalar-isoscalar
\begin{equation}
G\rightarrow G+G_0\text{ .} \label{shift}%
\end{equation}
where $G_0$ is the gluon condensate $G_0=\Lambda$. One can compute the branching ratios of the two- and three-body decay for the excited pseudoscalar glueball into scalar-pseudoscalar mesons and scalar glueball relative to the total decay width of the pseudoscalar glueball $\Gamma^{tot}_{\tilde{G}_2}$. \\

As another step, we consider the effective chiral Lagrangian that couples the excited pseudoscalar glueball field, $\tilde{G}$ to scalar and pseudoscalar mesons by the same means as the coupling of the pseudoscalar glueball to scalar and pseudoscalar quark-antiquark states as discussed in Ref.\cite{EshraimG}
\begin{equation}
\mathcal{L}_{\tilde{G}\Phi}^{int}=ic_{\tilde{G}\Phi}\tilde{G}\left(
\text{\textrm{det}}\Phi-\text{\textrm{det}}\Phi^{\dag}\right)  \text{ ,}
\label{intlag3}%
\end{equation}
where $c_{\tilde{G}\phi}$ is a dimensionless coupling constant. In this work we consider the case $N_f=4$ and the explicit representation of the scalar and pseudoscalar mesons reads \cite{EshraimC}
\begin{equation}\label{4}
\Phi=(S^{a}+iP^{a})t^{a}=\frac{1}{\sqrt{2}}
\left(%
\begin{array}{cccc}
  \frac{(\sigma_{N}+a^0_{0})+i(\eta_N +\pi^0)}{\sqrt{2}} & a^{+}_{0}+i \pi^{+} & K^{*+}_{0}+iK^{+} & D^{*0}_0+iD^0 \\
  a^{-}_{0}+i \pi^{-} & \frac{(\sigma_{N}-a^0_{0})+i(\eta_N -\pi^0)}{\sqrt{2}} & K^{*0}_{0}+iK^{0} & D^{*-}_0+iD^{-} \\
  K^{*-}_{0}+iK^{-} & \overline{K}^{*0}_{0}+i\overline{K}^{0} & \sigma_{S}+i\eta_{S} & D^{*-}_{S0}+iD^{-}_S\\
  \overline{D}^{*0}_0+i\overline{D}^0 & D^{*+}_0+iD^{+} & D^{*+}_{S0}+iD^{+}_S & \chi_{C0}+i\eta_C\\
\end{array}%
\right),
\end{equation}
The
multiplet $\Phi$ transforms as $\Phi\rightarrow U_{L}\Phi
U_{R}^{\dagger}$ under $U_{L}(4)\times U_{R}(4)$ chiral
transformations, whereas $U_{L(R)}=e^{-i\theta_{L(R)}^at^a}$ is an
element of $U(4)_{R(L)}$, under parity which
$\Phi(t,\overrightarrow{x})\rightarrow\Phi^{\dagger}(t,-\overrightarrow{x})$,
and under charge conjugation $\Phi\rightarrow\Phi^{\dagger}$. The
determinant of $\Phi$ is invariant under $SU(4)_{L} \times
SU(4)_{R}$, but not under $U(1)_{A}$ because ${\rm det
\Phi}\rightarrow {\rm det} U_{A}\Phi
U_A=e^{-i\theta_{A}^0\sqrt{2N_f}}{\rm det \Phi}\neq {\rm det
\Phi}$. The pseudoscalar glueball $\tilde{G}$ is invariant under
$U(4)_{L} \times U(4)_{R}$ transformations, under parity,
$\tilde{G}(t,\overrightarrow{x}) \rightarrow-
\tilde{G}(t,-\overrightarrow{x})$, and charge conjugation
$\tilde{G}\rightarrow \tilde{G}$. All this leads to the interaction
Lagrangian $\mathcal{L}_{\tilde{G}}^{int}$ of Eq.\ (\ref{intlag})
being invariant under $SU(4)_{L} \times SU(4)_{R}$, parity, and
charge conjugation. As before, Eq.\ (\ref{intlag}) is not invariant
under $U_{A}(1)$ .\\
The additional (pseudo)scalar charmed mesons appear in the fourth line and fourth column. In the scalar sector, open charmed meson $D^{\ast0,\pm}_0$ and strange charmed meson $D^{\ast\pm}_{S0}$ are assigned to $D^\ast_0(2400)^{0,\pm}$ and $D_{S0}^\ast(2317)^\pm$ \cite{EshraimC}, respectively. In the pseudoscalar sector there are an open charmed state $D^{0,\pm}$, open strange-charmed states $D_S^{\pm}$, and a hidden charmed ground state $\eta_{C}(1S)$. \\

In addition to shift the light scalar-isoscalar fields as seen in Eq. (\ref{shift}), one has to shift the charm-anticharm scalar field $\chi_{C0}$ by its vacuum expectation value $\phi_C$ to implement the spontaneous symmetry breaking as
\begin{equation}
\chi_{C0} \rightarrow \chi_{C0}+\phi_C\,,\label{scshift}
\end{equation}
where $\phi_C$ is the charm quark-antiquark condensates, which is fixed in the Ref. \cite{EshraimC}, as $\phi_C=176$ MeV.\\
To extend to the $N_f=4$ case, one adds to the shifting the axial-vector fields in Eq.(\ref{psz}) the following axial-vector charmonium state

\begin{equation}
\eta_C \rightarrow Z_{\eta_C}\,\eta_C\,,\label{vecshift}
\end{equation}
where the renormalization wave function is $Z_{\eta_C}=1.1189$ \cite{EshraimTH}. By including Eqs. (\ref{shift}, \ref{psz}, \ref{scshift}, \ref{vecshift}) in the Lagrangian (\ref{intlag3}), one obtains the relevant tree level vertices for the decay processes of the excited pseudoscalar glueball, $\tilde{G}$, as supplied in Appendix A4. The branching ratio for the decay of $\tilde{G}$  into two pions and one charmonium state $\eta_C$ is given as 
\begin{equation}
\Gamma_{\tilde{G}\rightarrow \eta_C\pi\pi}/\Gamma_{\tilde{G}_3}^{tot}=0.001 \,.
\end{equation}
This is of special interest, as it opens up the possibility for the decay of the excited pseudoscalar glueball into a charmonium state. 
The results of the branching ratios of $\tilde{G}$ for two- and three-body decays into states including scalar glueball and scalar-isoscalar, $f_0(1370),\,f_0(1500)$ and $f_0(1710)$, and (pseud)scalar states are reported in Table II and Table III, respectively, from the Lagrangian (\ref{intlag2}) and the Lagrangian (\ref{intlag3})

\begin{center}%
\begin{table}[h] \centering
\begin{tabular}
[c]{|c|c|c|c|}\hline
 Case (i):$\mathcal{L}_{\tilde{G}G}^{int}$ & The theoretical result & Case (ii):$\mathcal{L}_{\tilde{G}\Phi}^{int}$ & The theoretical result \\\hline
 $\Gamma_{\tilde{G}\rightarrow a_0\pi}/\Gamma_{\tilde{G}_2}^{tot}$ & $0.0325$ & $\Gamma_{\tilde{G}\rightarrow a_0\pi}/\Gamma_{\tilde{G}_3}^{tot}$ & $0.0313$ \\\hline
 $\Gamma_{\tilde{G}\rightarrow KK_S}/\Gamma_{\tilde{G}_2}^{tot}$ & $0.032$ & $\Gamma_{\tilde{G}\rightarrow KK_S}/\Gamma_{\tilde{G}_3}^{tot}$ & $0.001$\\\hline
 $\Gamma_{\tilde{G}\rightarrow \eta f_0(1370)}/\Gamma_{\tilde{G}_2}^{tot}$ & $0.00004$ & $\Gamma_{\tilde{G}\rightarrow \eta f_0(1370)}/\Gamma_{\tilde{G}_3}^{tot}$ & $0.0014$ \\\hline 
 $\Gamma_{\tilde{G}\rightarrow \eta' f_0(1370)}/\Gamma_{\tilde{G}_2}^{tot}$ & $0.048$ & $\Gamma_{\tilde{G}\rightarrow \eta' f_0(1370)}/\Gamma_{\tilde{G}_3}^{tot}$ & $0.031$\\\hline
 $\Gamma_{\tilde{G}\rightarrow \eta f_0(1500)}/\Gamma_{\tilde{G}_2}^{tot}$  & $0.0068$ & $\Gamma_{\tilde{G}\rightarrow \eta f_0(1500)}/\Gamma_{\tilde{G}_3}^{tot}$ & $0.0067$\\\hline
 $\Gamma_{\tilde{G}\rightarrow \eta' f_0(1500)}/\Gamma_{\tilde{G}_2}^{tot}$ & $0.0219$ & $\Gamma_{\tilde{G}\rightarrow \eta' f_0(1500)}/\Gamma_{\tilde{G}_3}^{tot}$ & $0.0214$\\\hline
 $\Gamma_{\tilde{G}\rightarrow \eta f_0(1710)}/\Gamma_{\tilde{G}_2}^{tot}$ & $0.0008$ & $\Gamma_{\tilde{G}\rightarrow \eta f_0(1710)}/\Gamma_{\tilde{G}_3}^{tot}$ & $0.0007$\\\hline
 $\Gamma_{\tilde{G}\rightarrow \eta' f_0(1710)}/\Gamma_{\tilde{G}_2}^{tot}$ & $0.001$  & $\Gamma_{\tilde{G}\rightarrow \eta' f_0(1710)}/\Gamma_{\tilde{G}_3}^{tot}$ & $0.001$\\\hline
\end{tabular}%
\caption{Branching ratios for the decays of the excited pseudoscalar glueball $\tilde
{G}$ into $PS$ and into $\eta$ and $\eta'$ and one of the scalar-isoscalar states; $f_0(1370),\,f_0(1500)$ and $f_0(1710)$ which correspond to the scalar glueball \cite{staniglueball}.}%
\end{table}%
\end{center}

\begin{center}%
\begin{table}[h] \centering
\begin{tabular}
[c]{|c|c|c|c|c|c|c|}\hline
 Case (i):$\mathcal{L}_{\tilde{G}G}^{int}$ & The theoretical result & Case (ii):$\mathcal{L}_{\tilde{G}\Phi}^{int}$ & The theoretical result \\\hline
$\Gamma_{\tilde{G}\rightarrow \eta \pi\pi}/\Gamma_{\tilde{G}_2}^{tot}$ & $0.095$ & $\Gamma_{\tilde{G}\rightarrow \eta \pi\pi}/\Gamma_{\tilde{G}_3}^{tot}$ & $0.1376$ \\\hline
$\Gamma_{\tilde{G}\rightarrow \eta' \pi\pi}/\Gamma_{\tilde{G}_2}^{tot}$ & $ 0.111$ &$\Gamma_{\tilde{G}\rightarrow \eta' \pi\pi}/\Gamma_{\tilde{G}_3}^{tot}$ & $0.1069$ \\\hline 
$\Gamma_{\tilde{G}\rightarrow a_0KK_S}/\Gamma_{\tilde{G}_2}^{tot}$ & $0.0026$ &  $\Gamma_{\tilde{G}\rightarrow a_0KK_S}/\Gamma_{\tilde{G}_3}^{tot}$ & $0.0025$ \\\hline
$\Gamma_{\tilde{G}\rightarrow \eta a_0 a_0}/\Gamma_{\tilde{G}_2}^{tot}$ & $0.0001$ & $\Gamma_{\tilde{G}\rightarrow \eta a_0 a_0}/\Gamma_{\tilde{G}_3}^{tot}$ & $0.0001$ \\\hline
$\Gamma_{\tilde{G}\rightarrow a_0\pi f_0(1370)}/\Gamma_{\tilde{G}_2}^{tot}$ & $0.0003$ &$\Gamma_{\tilde{G}\rightarrow a_0\pi f_0(1370)}/\Gamma_{\tilde{G}_3}^{tot}$ & $0.0003$ \\\hline
$\Gamma_{\tilde{G}\rightarrow a_0\pi f_0(1500)}/\Gamma_{\tilde{G}_2}^{tot}$ & $0.0034$ & $\Gamma_{\tilde{G}\rightarrow a_0\pi f_0(1500)}/\Gamma_{\tilde{G}_3}^{tot}$ & $0.0032$  \\\hline
$\Gamma_{\tilde{G}\rightarrow a_0\pi f_0(1710)}/\Gamma_{\tilde{G}_2}^{tot}$ & $0.0001$ &$\Gamma_{\tilde{G}\rightarrow a_0\pi f_0(1710)}/\Gamma_{\tilde{G}_3}^{tot}$ & $0.0001$ \\\hline
$\Gamma_{\tilde{G}\rightarrow \eta f^2_0(1370)}/\Gamma_{\tilde{G}_2}^{tot}$ & $0.0003$ &$\Gamma_{\tilde{G}\rightarrow \eta f^2_0(1370)}/\Gamma_{\tilde{G}_3}^{tot}$ & $0.001$ \\\hline
$\Gamma_{\tilde{G}\rightarrow \eta'f^2_0(1370)}/\Gamma_{\tilde{G}_2}^{tot}$ & $0.03\times 10^{-6}$ & $\Gamma_{\tilde{G}\rightarrow \eta'f^2_0(1370)}/\Gamma_{\tilde{G}_3}^{tot}$ & $0.006\times 10^{-6}$ \\\hline
$\Gamma_{\tilde{G}\rightarrow \eta f^2_0(1500)}/\Gamma_{\tilde{G}_2}^{tot}$ & $0.00004$ &$\Gamma_{\tilde{G}\rightarrow \eta f^2_0(1500)}/\Gamma_{\tilde{G}_3}^{tot}$ & $0.00001$ \\\hline
$\Gamma_{\tilde{G}\rightarrow \eta f_0(1370)f_0(1500)}/\Gamma_{\tilde{G}_2}^{tot}$ & $0.00003$ & $\Gamma_{\tilde{G}\rightarrow \eta f_0(1370)f_0(1500)}/\Gamma_{\tilde{G}_3}^{tot}$ & $0.0001$ \\\hline
$\Gamma_{\tilde{G}\rightarrow \eta f_0(1370)f_0(1710)}/\Gamma_{\tilde{G}_2}^{tot}$ & $3.798\times 10^{-6}$ &$\Gamma_{\tilde{G}\rightarrow \eta f_0(1370)f_0(1710)}/\Gamma_{\tilde{G}_3}^{tot}$ & $7.25\times 10^{-6}$  \\\hline
$\Gamma_{\tilde{G}\rightarrow K K_S f_0(1370)}/\Gamma_{\tilde{G}_2}^{tot}$ & $0.0025$ & $\Gamma_{\tilde{G}\rightarrow K K_S f_0(1370)}/\Gamma_{\tilde{G}_3}^{tot}$ & $0.0025$ \\\hline
$\Gamma_{\tilde{G}\rightarrow K K_S f_0(1500)}/\Gamma_{\tilde{G}_2}^{tot}$ &$0.00013$ & $\Gamma_{\tilde{G}\rightarrow K K_S f_0(1500)}/\Gamma_{\tilde{G}_3}^{tot}$ & $0.00013$  \\\hline
$\Gamma_{\tilde{G}\rightarrow K K_S f_0(1710)}/\Gamma_{\tilde{G}_2}^{tot}$ &$6.2\times 10^{-6}$ & $\Gamma_{\tilde{G}\rightarrow K K_S f_0(1710)}/\Gamma_{\tilde{G}_3}^{tot}$ & $4.75\times 10^{-6}$ \\\hline
$\Gamma_{\tilde{G}\rightarrow KK\eta}/\Gamma_{\tilde{G}_2}^{tot}$ &$0.0668$ &$\Gamma_{\tilde{G}\rightarrow KK\eta}/\Gamma_{\tilde{G}_3}^{tot}$ & $0.0643$ \\\hline
$\Gamma_{\tilde{G}\rightarrow KK\eta'}/\Gamma_{\tilde{G}_2}^{tot}$ &$0.045$ &$\Gamma_{\tilde{G}\rightarrow KK\eta'}/\Gamma_{\tilde{G}_3}^{tot}$ & $0.044$ \\\hline
$\Gamma_{\tilde{G}\rightarrow K_SK_S\eta}/\Gamma_{\tilde{G}_2}^{tot}$ &$0.0002$& $\Gamma_{\tilde{G}\rightarrow K_SK_S\eta}/\Gamma_{\tilde{G}_3}^{tot}$ & $0.0002$ \\\hline
$\Gamma_{\tilde{G}\rightarrow \eta^3}/\Gamma_{\tilde{G}_2}^{tot}$ & $0.024$&$\Gamma_{\tilde{G}\rightarrow \eta^3}/\Gamma_{\tilde{G}_3}^{tot}$ & $0.0233$ \\\hline
$\Gamma_{\tilde{G}\rightarrow \eta'^3}/\Gamma_{\tilde{G}_2}^{tot}$ & $0.0048$ &$\Gamma_{\tilde{G}\rightarrow \eta'^3}/\Gamma_{\tilde{G}_3}^{tot}$ & $0.0046$ \\\hline
$\Gamma_{\tilde{G}\rightarrow \eta'\eta^2}/\Gamma_{\tilde{G}_2}^{tot}$ &$ 0.005$& $\Gamma_{\tilde{G}\rightarrow \eta'\eta^2}/\Gamma_{\tilde{G}_3}^{tot}$ & $0.0048$  \\\hline
$\Gamma_{\tilde{G}\rightarrow \eta'^2\eta}/\Gamma_{\tilde{G}_2}^{tot}$ & $0.0035$ &$\Gamma_{\tilde{G}\rightarrow \eta'^2\eta}/\Gamma_{\tilde{G}_3}^{tot}$ & $0.0034$ \\\hline
$\Gamma_{\tilde{G}\rightarrow KK\pi}/\Gamma_{\tilde{G}_2}^{tot}$ & $0.489$ & $\Gamma_{\tilde{G}\rightarrow KK\pi}/\Gamma_{\tilde{G}_3}^{tot}$ & $0.471$  \\\hline
$\Gamma_{\tilde{G}\rightarrow K_SK_S\pi}/\Gamma_{\tilde{G}_2}^{tot}$ & $0.002$ &$\Gamma_{\tilde{G}\rightarrow K_SK_S\pi}/\Gamma_{\tilde{G}_3}^{tot}$ & $0.0057$ \\\hline
\end{tabular}%
\caption{Branching ratios for the decays of the excited pseudoscalar glueball $\tilde
{G}$ into the scalar-isoscalar states and (pseudo)scalar mesons.}%
\end{table}%
\end{center}

Tables II and III show the excited pseudoscalar glueball decays into scalar-isoscalar states, $f_0(1370),\,f_0(1500)$ and $f_0(1710)$, by including the full mixing pattern above $1$ GeV and $\tilde{G}$ decay into the scalar glueball which corresponds to  the resonance $f_0(1710)$ \cite{staniglueball}. Furthermore, the results for $\mathcal{L}_{\tilde{G}G}^{int}$ and $\mathcal{L}_{\tilde{G}\Phi}^{int}$ are very close in the two- and three-body decays, which could provide valuable insight for experiment.  

\newpage

\section{Conclusion}

In this work we have presented three chirally invariant effective Lagrangians. The first one describes the interaction of the excited pseudoscalar glueball with the lowest pseudoscalar glueball and (pseudo)scalar mesons, for the three-flavour case $N_f=3$. We have studied the three-body decays of the excited pseudoscalar glueball with a mass of $3.7$ GeV, including decays into one pseudoscalar glueball with a mass of $2.6$ GeV and two pseudoscalar mesons $\Gamma_{\tilde{G}\rightarrow \tilde{G}'PP}$. The second Lagrangian describes the interaction of the excited pseudoscalar glueball with a scalar glueball and (pseudo)scalar mesons in the case of $N_f=3$. From this effective Lagrangian, we have computed the decays of the excited pseudoscalar glueball, also with a reference mass of $3.7$ GeV, into two- and three- (pseudo)scalar mesons and scalar-isoscalar states $f_0(1370),\,f_0(1500)$ and $f_0(1710)$, where the resonance $f_0=(1710)$ is identified with the scalar glueball. The third chiral Lagrangian extends treatment to the four-flavour case ($N_f=4$) including charmonium states. This study yields an interesting result for the decay of the excited pseudoscalar glueball into the charmonium state $\eta_C$ as seen in $\Gamma_{\tilde{G}\rightarrow \eta_C\pi\pi}$. Furthermore, from the third effective Lagrangian we have computed  the two and three-body decays for the excited pseudoscalar glueball into (pseudo)scalar mesons and the scalar-isoscalar states.\\
\indent We have presented the results as branching ratios to eliminate the unknown overall normalization. We conclude that the excited pseudoscalar glueball with a mass of about $3.7$ GeV may decay into the pseudoscalar glueball with a mass of $2.6$ GeV, the charmonium state $\eta_C$, the scalar glueball and the (pseudo)scalar mesons with clearly defined branching ratios. 
The resulting numbers can serve as a guide for the BESIII and for the corresponding upcoming experiments with the PANDA detector at FAIR.

\section*{Acknowledgements}

The authors thank Dirk Rischke and Horst Stoecker for useful discussions. Financial support from Female program for HIC for FAIR is acknowledged.

\appendix

\section{Details of the calculation}

\subsection{The full mesonic Lagrangian}

\label{app1}

The chirally invariant $U(N_{f})_{L}\times U(N_{f})_{R}$ Lagrangian for the
low-lying mesonic states with (pseudo)scalar and (axial-)vector quantum
numbers has the form
\begin{align}
\mathcal{L}_{mes}  &  =\mathrm{Tr}[(D_{\mu}\Phi)^{\dagger}(D_{\mu}\Phi
)]-m_{0}^{2}\mathrm{Tr}(\Phi^{\dagger}\Phi)-\lambda_{1}[\mathrm{Tr}%
(\Phi^{\dagger}\Phi)]^{2}-\lambda_{2}\mathrm{Tr}(\Phi^{\dagger}\Phi
)^{2}\nonumber\\
&  -\frac{1}{4}\mathrm{Tr}[(L^{\mu\nu})^{2}+(R^{\mu\nu})^{2}]+\mathrm{Tr}%
[(\frac{m_{1}^{2}}{2}+\Delta)(L_{\mu}^{2}+R_{\mu}^{2})]+\mathrm{Tr}%
[H(\Phi+\Phi^{\dagger})]\nonumber\\
&  +c_{1}(\mathrm{det}\Phi-\mathrm{det}\Phi^{\dagger})^{2}+i\frac{g_{2}}%
{2}\{\mathrm{Tr}(L_{\mu\nu}[L^{\mu},L^{\nu}])+\mathrm{Tr}(R_{\mu\nu}[R^{\mu
},R^{\nu}])\}\nonumber\\
&  +\frac{h_{1}}{2}\mathrm{Tr}(\Phi^{\dagger}\Phi)\mathrm{Tr}\left(  L_{\mu
}^{2}+R_{\mu}^{2}\right)  +h_{2}\mathrm{Tr}[\left\vert L_{\mu}\Phi\right\vert
^{2}+\left\vert \Phi R_{\mu}\right\vert ^{2}]\nonumber\\
&  +2h_{3}\mathrm{Tr}(L_{\mu}\Phi R^{\mu}\Phi^{\dagger}). \label{fulllag}%
\end{align}
where
\begin{equation}\label{4}
L^\mu=(V^a+i\,A^a)^{\mu}\,t^a=\frac{1}{\sqrt{2}}
\left(%
\begin{array}{cccc}
  \frac{\omega_N+\rho^{0}}{\sqrt{2}}+ \frac{f_{1N}+a_1^{0}}{\sqrt{2}} & \rho^{+}+a^{+}_1 & K^{*+}+K^{+}_1 & D^{*0}+D^{0}_1 \\
  \rho^{-}+ a^{-}_1 &  \frac{\omega_N-\rho^{0}}{\sqrt{2}}+ \frac{f_{1N}-a_1^{0}}{\sqrt{2}} & K^{*0}+K^{0}_1 & D^{*-}+D^{-}_1 \\
  K^{*-}+K^{-}_1 & \overline{K}^{*0}+\overline{K}^{0}_1 & \omega_{S}+f_{1S} & D^{*-}_{S}+D^{-}_{S1}\\
  \overline{D}^{*0}+\overline{D}^{0}_1 & D^{*+}+D^{+}_1 & D^{*+}_{S}+D^{+}_{S1} & J/\psi+\chi_{C1}\\
\end{array}%
\right)^\mu,
\end{equation}
and
\begin{equation}\label{5}
R^\mu=(V^a-i\,A^a)^\mu\,t^a=\frac{1}{\sqrt{2}}
\left(%
\begin{array}{cccc}
  \frac{\omega_N+\rho^{0}}{\sqrt{2}}- \frac{f_{1N}+a_1^{0}}{\sqrt{2}} & \rho^{+}-a^{+}_1 & K^{*+}-K^{+}_1 & D^{*0}-D^{0}_1 \\
  \rho^{-}- a^{-}_1 &  \frac{\omega_N-\rho^{0}}{\sqrt{2}}-\frac{f_{1N}-a_1^{0}}{\sqrt{2}} & K^{*0}-K^{0}_1 & D^{*-}-D^{-}_1 \\
  K^{*-}-K^{-}_1 & \overline{K}^{*0}-\overline{K}^{0}_1 & \omega_{S}-f_{1S} & D^{*-}_{S}-D^{-}_{S1}\\
  \overline{D}^{*0}-\overline{D}^{0}_1 & D^{*+}-D^{+}_1 & D^{*+}_{S}-D^{+}_{S1} & J/\psi-\chi_{C1}\\
\end{array}%
\right)^\mu.
\end{equation}
The fields $\omega_N,\,\omega_S,\, \overrightarrow{\rho}, \,f_{1N}, \,f_{1S},\, \overrightarrow{a_1},\, K^*,\, K^+_0$ and $K_1$ are assigned to the light physical resonances $\omega(782), \phi(1020),\, \rho(770),\, f_1(1420),\, a_1(1260),\,K^*(892),\,K^*_0(1430)$, and $K_1(1270)$ [or $K_1(1400)$, see the discussion in Refs. \cite{dick, LutzK1}] mesons, respectively. The charmed fields $D^{*0},\, D^*, \, \chi_{C1},\, J/\psi,$ and $D_{S1}$ are assigned to heavy physical resonsnces $D^*(2007)^0,\, D^+(2010)^{\pm},\, \chi_{C1}(1P),\,J/\psi(1S)$, and $D_{S1}(2536)$, respectively \cite{EshraimC, EshraimTH}.\\

In the present context we are interested in the wave-function renormalization
constants $Z_{i}$ introduced in Eq.\ (\ref{psz}). Their explicit expressions
read \cite{EshraimTH}:%
\begin{equation}
Z_{\pi}=Z_{\eta_{N}}=\frac{m_{a_{1}}}{\sqrt{m_{a_{1}}^{2}-g_{1}^{2}\phi
_{N}^{2}}}\;, \label{zpi}%
\end{equation}%
\begin{equation}
Z_{K}=\frac{2m_{K_{1}}}{\sqrt{4m_{K_{1}}^{2}-g_{1}^{2}(\phi_{N}+\sqrt{2}%
\phi_{S})^{2}}}\;,
\end{equation}%
\begin{equation}
Z_{K_{S}}=\frac{2m_{K^{\star}}}{\sqrt{4m_{K^{\star}}^{2}-g_{1}^{2}(\phi
_{N}-\sqrt{2}\phi_{S})^{2}}}\;,
\end{equation}%
\begin{equation}
Z_{\eta_{S}}=\frac{m_{f_{1S}}}{\sqrt{m_{f_{1S}}^{2}-2g_{1}^{2}\phi_{S}^{2}}%
}\;. \label{zets}%
\end{equation}
\begin{align}
Z_{\eta_{C}}=\frac{m_{\chi_{C1}}}{\sqrt{m_{\chi_{C1}}^{2}-2g_{1}^{2}\phi
_{C}^{2}}}\,,
\end{align}

\subsection{Explicit form of the Lagrangian in Eq.\ (\ref{intlag})}

\label{app2}

After performing the field transformations in Eqs.\ (\ref{shift}) and
(\ref{psz}), the effective Lagrangian (\ref{intlag}) takes the form:
\begin{align}\label{exL1}
\mathcal{L}_{\tilde{G}\tilde{G}'}^{int}  &  =\frac{1}{2}c_{\tilde{G}\tilde{G}'}%
\tilde{G}\tilde{G}'(a_0^0a_0^0+2a_0^-a_0^++2Z_K^2K^0\overline{K}^{0}+2Z_K^2K^-K^++ 2Z^2_{K_{S}}K_{S}^{0}\overline{K}_S^{0}+2Z_{K_{S}}K_{S}^{-}K_S^+\nonumber\\
&  Z_{\eta_{N}}^2\eta_{N}^2+Z_{\eta_{S}}^2\eta_{S}^2+Z_{\pi}^2\pi^{2}_0+2Z_{\pi}^2\pi^{-}\pi^{+}+2\sigma
_{S}^2+\sigma_{N}^2+2\sqrt{2}\phi_N \sigma_{N}+2\sigma_S\phi_S+\phi_N^2+\phi_{S}^2)\text{ .}%
\end{align}

Note that, some decay channels of the excited pseudoscalar glueball, $\tilde{G}$, are not kinematically allowed, because the mass of the decaying particle is larger than the summation mass of the decay products $M<\sum_i^3 m_i$, which is summarized as follows
\begin{align}
\Gamma_{\tilde{G}\rightarrow \tilde{G}' a_0 a_0}=0,\,\,\,\,\,\, &\Gamma_{\tilde{G}\rightarrow \tilde{G}' K_SK_S}=0\,,\\
\Gamma_{\tilde{G}\rightarrow \tilde{G}' \sigma_N}=0,\,\,\,\,\,\, &\Gamma_{\tilde{G}\rightarrow \tilde{G}' \sigma_S}=0\,,\\
\Gamma_{\tilde{G}\rightarrow \tilde{G}' \sigma_N^2}=0,\,\,\,\,\,\, &\Gamma_{\tilde{G}\rightarrow \tilde{G}' \sigma_S^2}=0\,.
\end{align}

There is a mixing between the excited pseudoscalar glueball, $\tilde{G}$, and the pseudoscalar glueball, $\tilde{G}'$, appear in the the Lagrangian (\ref{exL1}) in the term $\frac{1}{2}c_{\tilde{G}\tilde{G}'}%
\tilde{G}\tilde{G}'(\phi_N+\phi_S)$. The full $\tilde{G}-\tilde{G}'$ interaction Lagrangian has the form

\begin{equation}\label{70}
 \mathcal{L}_{\widetilde{G},\,\eta_C}= \frac{1}{2}(\partial_\mu\widetilde{G})^2+\frac{1}{2}(\partial_\mu\widetilde{G}')^2-\frac{1}{2}m_{\widetilde{G}}^2\widetilde{G}^2-\frac{1}{2}m_{\widetilde{G}'}^2\widetilde{G}'^2+Z_{\widetilde{G}\widetilde{G}'}\widetilde{G}\,\widetilde{G}'\,,
\end{equation}
where
\begin{equation}\label{70}
 Z_{\widetilde{G}\widetilde{G}'}=\frac{1}{2}\,c_{\tilde{G}\tilde{G}'}%
\tilde{G}\tilde{G}'(\phi_N+\phi_S)\,.
\end{equation}
The physical fields $\widetilde{G}$ and $\widetilde{G}'$ can be obtained
through an SO(2) rotation
\begin{equation}\label{8}
\left(%
\begin{array}{c}
 \widetilde{G}_1 \\
 \widetilde{G}'_1 \\
\end{array}%
\right)=\left(%
\begin{array}{cccc}
 cos\phi & sin\phi  \\
 -sin\phi & cos\phi \\
\end{array}%
\right)=\left(%
\begin{array}{c}
 \widetilde{G}\\
 \widetilde{G}' \\
\end{array}%
\right),
\end{equation}
with
\begin{equation}\label{70}
m^2_{\widetilde{G}_1}=m^2_{\widetilde{G}'}\,
sin^2\phi+m^2_{\widetilde{G}}cos^2\phi-Z_{\widetilde{G}\widetilde{G}'}sin(2\phi),
\end{equation}
\begin{equation}\label{70}
m^2_{\widetilde{G}'_1}=m^2_{\widetilde{G}}\,
sin^2\phi+m^2_{\widetilde{G}'}cos^2\phi+Z_{\widetilde{G}\widetilde{G}'}sin(2\phi),
\end{equation}
where the mixing angle $\phi$ reads
\begin{equation}\label{70}
\phi=\frac{1}{2}arctan\bigg[\frac{c_{\widetilde{G}\widetilde{G}'}\,(\phi_N^2+\phi_S^2)}{(m^2_{\widetilde{G}}-m^2_{\widetilde{G}'})}\bigg]
\end{equation}
where $c_{\widetilde{G}\widetilde{G}'}$ is a dimensionless coupling constant
between $\widetilde{G}\widetilde{G}'$.

\subsection{Explicit form of the Lagrangian in Eq.\ (\ref{intlag2})}

\label{app3}

After executing the field transformations in Eqs.\ (\ref{shift}) and
(\ref{psz}), the chiral effective Lagrangian (\ref{intlag2}) takes the form:

\begin{align}\label{exL2}
\mathcal{L}_{\tilde{G}\Phi}  &  =\frac{1}{4}\phi_{C0}c_{\tilde{G}\Phi}
\tilde{G}\big[-2Z_KZ_{K_S}a^-_0\overline{K}^0_SK^+-2Z_KZ_{K_S}a^-_0\overline{K}^0 K^+_S-\sqrt{2}Z_K^2Z_{\eta_N}\overline{K}^0K^0\eta_N-\sqrt{2}Z_K^2Z_{\eta_N}K^-K^+\eta_N\\ \nonumber
&+\sqrt{2}Z_{K_S}^2 Z_{\eta_N} \overline{K}^0_S K^0_S\eta_N+\sqrt{2}Z_{K_S}^2Z_{\eta_N}K^-_S K^+_S\eta_N+Z_{\eta_S} a^0_0 \eta_S +Z_{\eta_N}^2Z_{\eta_S}\eta_N^2\eta_S-\sqrt{2}Z_\pi Z_{K}^2\pi^0\overline{K}^0K^0\\\nonumber
&+\sqrt{2}Z_\pi Z_K^2 \pi^0K^-K^++\sqrt{2}Z_\pi Z_{K_S}^2\pi^0\overline{K}_S^0K_S^0-\sqrt{2}Z_\pi Z_{K_S}^2 \pi^0 K_S^-K^+_S-Z_{\pi}^2Z_{\eta_S}\pi^0\pi^0\eta_S+2Z_\pi Z_{K}^2\pi^-\overline{K}^0K^+\\\nonumber
&-2Z_\pi Z_{K_S}^2\pi^-\overline{K}^0_SK^+_S+2Z_\pi Z_{K}^2 \pi^+ K^0K^--2Z_\pi Z_{K_S}^2 \pi^+ K^0_SK^-_S-2Z_\pi^2Z_{\eta_S} \pi^-\pi^+\eta_S+2Z_KZ_{K_S} \overline{K}^0K^0_S\sigma_N\\\nonumber
&+2Z_KZ_{K_S}\overline{K}^0_SK^0\sigma_N+2Z_KZ_{K_S}\overline{K}^-_SK^+\sigma_N+2Z_KZ_{K_S}K^-K^+_S\sigma_N-2Z_{\eta_S}\eta_S\sigma_N^2+2Z_\pi \pi^+ a^-_0\sigma_S\\\nonumber
&-2\sqrt{2}Z_{\eta_N}\eta_N\sigma_N\sigma_S+\sqrt{2}Z_KZ_{K_S}\Phi_N \overline{K}^0K^0_S+\sqrt{2}Z_KZ_{K_S}\Phi_N \overline{K}^0_SK^0+\sqrt{2}Z_KZ_{K_S}\Phi_N K^-_SK^+\\\nonumber
&+\sqrt{2}Z_KZ_{K_S}\Phi_N K^-K^+_S-2\sqrt{2}Z_{\eta_S}\Phi_N\eta_S\sigma_N-2Z_{\eta_N}\Phi_N\eta_N\sigma_S-Z_{\eta_S}\Phi_N^2\eta_S+2Z_\pi\Phi_S\pi^+a_0^--2\sqrt{2}Z_{\eta_N}\Phi_S\eta_N\sigma_N\\\nonumber
&-2Z_{eta_N}\Phi_N\Phi_S\eta_N+a_0^0(\sqrt{2}Z_KZ_{K_S}\overline{K}^0K^0_S+\sqrt{2}Z_K Z_{K_S} \overline{K}^0_SK^0-\sqrt{2}Z_K Z_{K_S}+2Z_\pi\pi^0\sigma_S+2Z_\pi\Phi_S\pi^0)\\\nonumber
&+2a^+_0(Z_K Z_{K_S}K^0_SK^--Z_K Z_{K_S}K^-_SK^0+Z_{\eta_S}a^-_0\eta_S+Z_\pi\pi^-\sigma_S+Z_\pi\Phi_S\pi^-)
\big]\text{ .}%
\end{align}

\subsection{Explicit form of the Lagrangian in Eq.\ (\ref{intlag3})}

\label{app4}

The corresponding interaction Lagrangian
from Eq.(\ref{intlag3}) (only the particles produced in tables II and III) is obtained by executing the field transformations in Eqs.\ (\ref{shift}), 
(\ref{psz}), (\ref{scshift}) and (\ref{vecshift}) as
\begin{align}\label{exL3}
\mathcal{L}_{\tilde{G}\Phi}  &  =\frac{1}{4}\phi_{C0}c_{\tilde{G}\Phi}
\tilde{G}\big\{-Z_\pi^2 Z_{\eta_S}\eta_S(\pi^0\pi^0+2\pi^-\pi^+)+Z_{\eta_S}(a_0^0a_0^0+2a_0^-a_0^+)\eta_S\\ \nonumber
&-Z_K Z_{K_S}[2a_0^+(K_S^0K^-+K^0K_S^-)+2a_0^-(\overline{K}_S^0K^++\overline{K}^0K_S^+)-\sqrt{2}a_0^0(K^0_S\overline{K}^0+K^0\overline{K}^0_S-K_S^-K^++K^-K_S^+)]\\ \nonumber
&+Z_\pi[(\pi^0a_0^0+\pi^+a_0^-+\pi^-a_0^+)(\sigma_S+\phi_S)]-\sqrt{2}Z_K^2Z_{\eta_N}\eta_N(K^0\overline{K}^0+K^-K^+)+\sqrt{2}Z_{K_S}^2Z_{\eta_N}\eta_N(K^0_S\overline{K}^0_S+K_S^-K^+_S)\\ \nonumber
&+Z_{\eta_N}^2Z_{\eta_S}\eta_N^2\,\eta_S+Z_\pi Z_{K}^2[\sqrt{2}(-K^0\overline{K}^0+K^-K^+)\pi^0+2(\overline{K}^0K^+\pi^-+K^0K^+\pi^0)]\\ \nonumber
&+Z_\pi Z_{K_S}^2[\sqrt{2}(K^0_S\overline{K}^0_S-K^-_SK^+_S)\pi^0-2(\overline{K}^0_SK^+_S\pi^-+K^0_SK^-_S\pi^+)]\\ \nonumber &
+2 Z_K Z_{K_S}(\overline{K}^0_S K^0+K^0_S\overline{K^0}+K^-_SK^++K_S^+K^-)\sigma_N+\sqrt{2} Z_K Z_{K_S}\phi_N(\overline{K}^0_S K^0+K^0_S\overline{K^0}+K^-_SK^++K_S^+K^-)\\ \nonumber
&-2Z_{\eta_S}\eta_S\sigma_N^2-2\sqrt{2}Z_{\eta_N}\eta_N\sigma_N\sigma_S-2\sqrt{2}Z_{\eta_S}\phi_N\eta_S\sigma_N-2Z_{\eta_N}\phi_N\eta_N\sigma_S-2\sqrt{2}Z_{\eta_N}\phi_S\eta_N\sigma_N
\big\}\text{ .}%
\end{align}

\subsection{Two-body decay}

The general formula of the two-body
decay width \cite{EshraimTH} is

\begin{equation}
\label{B1}\Gamma_{A\rightarrow BC}=\frac{S_{A\rightarrow BC}k(m_{A}%
,\,m_{B},\,m_{C})}{8 \pi m_{A}^{2}}|\mathcal{M}_{A\rightarrow
BC}|^{2},
\end{equation}

where A is the decaying particle, B and C are the decay products, $k(m_{A}%
,\,m_{B},\,m_{C})$ is the center-of-mass momentum of the two
particles produced in the decay, described as follows
\begin{equation}
k(m_{A},\,m_{B},\,m_{C})=\frac{1}{2m_{A}}\sqrt{m_{A}^{4}+(m_{B}^{2}-m_{C}%
^{2})^{2}-2m_{A}^{2}\,(m_{B}^{2}+m_{C}^{2})}\theta(m_{A}-m_{B}-m_{C}),
\label{B2}%
\end{equation}
$\mathcal{M}_{A\rightarrow BC}$ is the corresponding tree-level
decay amplitude, and $S_{A\rightarrow BC}$ refers to a
symmetrization factor (it equals $1$ if B and C are different and
it equals $1/2$ for two identical particles in the final state).

\subsection{Three-body decay}

\label{app3}

For completeness we report the explicit expression for the three-body decay
width for the process $\tilde{G}\rightarrow P_{1}P_{2}P_{3}$ \cite{PDG}:%
\[
\Gamma_{\tilde{G}\rightarrow P_{1}P_{2}P_{3}}=\frac{s_{\tilde{G}\rightarrow
P_{1}P_{2}P_{3}}}{32(2\pi)^{3}M_{\tilde{G}}^{3}}\int_{(m_{1}+m_{2})^{2}%
}^{(M_{\tilde{G}}-m_{3})^{2}}dm_{12}^{2}\int_{(m_{23})_{\min}}^{(m_{23}%
)_{\max}}|-i\mathcal{M}_{\tilde{G}\rightarrow P_{1}P_{2}P_{3}}|^{2}dm_{23}^{2}%
\]
where
\begin{align}
(m_{23})_{\min}  &  =(E_{2}^{\ast}+E_{3}^{\ast})^{2}-\left(  \sqrt{E_{2}%
^{\ast2}-m_{2}^{2}}+\sqrt{E_{3}^{\ast2}-m_{3}^{2}}\right)  ^{2}\text{ ,}\\
(m_{23})_{\max}  &  =(E_{2}^{\ast}+E_{3}^{\ast})^{2}-\left(  \sqrt{E_{2}%
^{\ast2}-m_{2}^{2}}-\sqrt{E_{3}^{\ast2}-m_{3}^{2}}\right)  ^{2}\text{ ,}%
\end{align}
and%
\begin{equation}
E_{2}^{\ast}=\frac{m_{12}^{2}-m_{1}^{2}+m_{2}^{2}}{2m_{12}}\text{ , }%
E_{3}^{\ast}=\frac{M_{\tilde{G}}^{2}-m_{12}^{2}-m_{3}^{2}}{2m_{12}}\text{ .}%
\end{equation}
The quantities $m_{1},$ $m_{2},$ $m_{3}$ refer to the masses of the three
pseudoscalar states $P_{1},$ $P_{2}$, and $P_{3},$ $\mathcal{M}_{\tilde
{G}\rightarrow P_{1}P_{2}P_{3}}$ is the corresponding tree-level decay
amplitude, and $s_{\tilde{G}\rightarrow P_{1}P_{2}P_{3}}$ is a symmetrization
factor (it equals $1$ if all $P_{1},$ $P_{2}$, and $P_{3}$ are different, it
equals $2$ for two identical particles in the final state, and it equals $6$
for three identical particles in the final state).

\end{document}